\documentclass{article}
\usepackage{bm}
\usepackage{graphicx}

\newcommand{\op}[1]{\widehat{#1}}
\newcommand{\dagop}[1]{\widehat{#1}{}^{\dagger}}
\newcommand{\mc}[1]{\mathcal{#1}}

\begin{document}
\markboth{Drummond,  Deuar, Corney \& Vaughan}{Quantum dynamics in
phase-space}

\title{Quantum dynamics in phase space: From coherent states to the Gaussian
representation}

\author{P. D. Drummond$^{(1)}$, P. Deuar$^{(2)\dagger}$,  T. G. Vaughan$^{(1)}$
and J. F. Corney$^{(1)}$ \\
 $^{(1)}$ARC Centre of Excellence for Quantum-Atom Optics, \\
 School of Physical Sciences, University of Queensland, \\
Brisbane, QLD 4072, Australia\\
$^{(2)}$Van der Waals-Zeeman Instituut, Universiteit van Amsterdam,
\\
1018 XE Amsterdam, Netherlands\\
$^{\dagger}$Present address: LPTMS, B\^{a}t. 100, Universit\'{e} Paris-Sud 91405 Orsay cedex, France}

\maketitle
\begin{abstract}
We give an outlook on the future of coherence theory and many-body
quantum dynamics as experiments develop in the arena of ultra-cold
atoms. Novel results on quantum heating of center-of-mass temperature
in evaporative cooling and simulation methods for long-range interactions
are obtained, using positive-P phase-space techniques. 
\end{abstract}

\section{Coherence theory in the 21st century}

One recognition of important developments in coherence theory was
the $2005$ Nobel award in Physics, one half to Roy J. Glauber, `\textsl{for
his contribution to the} \textsl{quantum theory of} \textsl{optical coherence'}
, and one half to Ted Haensch and Jan Hall `\textsl{for their contributions
to the development of laser-based precision spectroscopy}' . This
richly deserved award recognizes crucial developments in quantum optics
and laser science in the second half of the twentieth century. One
may ask now\textbf{:} \textbf{\textsl{What is the future of coherence
theory?}}

One answer to this question lies in the groundbreaking work of experimentalists
working with ultra-cold atoms. Perhaps the ideal quantum system for
experimental investigation, ultracold atoms display many useful properties
under active investigation, including:

\begin{itemize}
\item Ultra-low temperatures to below 1nK 
\item Bose-Einstein condensates (BEC): atom `photons' 
\item Quantum superfluid degenerate Fermi gases (DFG): atom `electrons' 
\item `Superchemistry': stimulated bosonic molecule formation 
\item Atom lasers, atomic diffraction, atom interferometers 
\item Direct detection of atom coherence and correlations 
\end{itemize}
A crucial, common property of photons and ultracold gases is their
\textsl{simplicity} as many-body systems. This makes them ideal candidates
for both theoretical and experimental investigation in fundamental
science. The underlying interactions are well understood, the experimental
systems can be easily characterized by a few parameters, and interaction
strengths can be tuned.

Under these conditions, well-known theoretical models can be used
to high accuracy, thus combining ideas from coherence and many-body
theory. As well as being able to test and understand theories like
the Hubbard model, one has the possibility of new technologies of
unprecedented accuracy and subtlety. This is likely to lead to new
tests of macroscopic quantum mechanics and quantum superpositions,
which is undoubtedly one of the grand challenges of modern physics.

In this paper, we give a brief overview of recent directions that
coherence theory has taken since it originated in the quantum optics
area, as well as giving new theoretical results on examples of quantum
dynamics. First we describe some of the recent experimental developments
in quantum-atom optics, in which the role of correlations are becoming
increasingly important. Second, we review theoretical developments
in which coherence theory is being utilized to give new simulation
techniques that can handle the fundamental issue of quantum dynamics
of many-body systems.

\section{Quantum dynamical experiments}

As one of the new types of experiment on atomic coherence and correlations,
many laboratories are now able to carry out intrinsically \textsl{dynamical}
experiments on many-body systems, rather than the near-equilibrium
experiments of condensed matter physics. Dynamical results provide
a new probe into the properties of many-body systems. Clearly, the
future of coherence theory must include an understanding of how to
quantitatively predict the results of experiments involving the dynamical
evolution of many-body quantum systems far from thermal equilibrium.
Due to the rapid growth of ultra-cold atom facilities, these types
of experiment are now carried out in many laboratories. They test
quantum theory in regimes of large particle number, as they typically
involve $10^{2} - 10^{7}$ interacting particles, at temperatures
of around $100$nK. 

\begin{figure}
\begin{centering}\includegraphics[width=8cm]{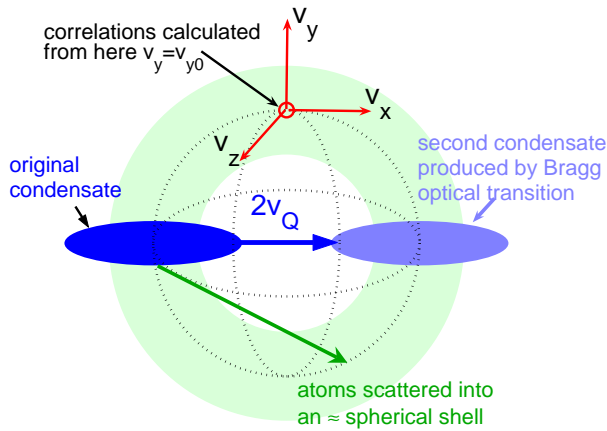} \par\end{centering}

\caption{Momentum-Space Diagram of an ultra-cold collision experiment.\label{fig:Momentum-Space-Diagram-of}}
\end{figure}

A schematic diagram of the type of experiment that we will focus on
is shown in Fig (\ref{fig:Momentum-Space-Diagram-of}). A condensate
is prepared with each atom in a quantum superposition of two different
momenta, resulting in effectively two condensates in relative motion,
occupying the same physical space. Subsequently, a strong scattering
commences, in which both individual particle and coherent many-body
effects play important roles.

In the remainder of this section, we will outline two recent experiments
of this type, in which many-body effects clearly play a role. There
are other experiments as well, and indeed this is a very rapidly developing
field.

\subsection{Collisions in Sodium BEC}

An example of quantum dynamics is provided by a series of experiments
by the Ketterle group at MIT\cite{Vogels2002} that involve the three-dimensional
collision of two sodium (Na) BECs, as in the schematic diagram of
Fig (\ref{fig:Momentum-Space-Diagram-of}). The two BECs are produced
from a single initially ellipsoidal trapped cloud, as found in a non-spherical
magnetic trap.

The initial condensate is then split into two halves, with a large
relative velocity, and the trap is turned off. This leaves two condensates
that are spatially overlapped and have a relative velocity, so that
a collision occurs. Also observed in these experiments are: amplification
of seed pulses during a collision, interaction of condensates with
lattices, and quantum reflection from a mirror. These condensates
typically include $10^{7}$ or more interacting bosons, and measurements
usually involve the observation of density distributions.

\subsection{Metastable Helium experiments}

A recent experimental configuration that allows the retrieval of much
more information is the interaction of metastable Helium condensates.
These have a distinct advantage over the alkali metals like sodium
or rubidium, in that single atom arrivals at a detector multi-channel
plate can be readily detected, owing to the high excitation energy
(around $20$eV) of the metastable atoms. This allows atomic correlations
to be measured directly\cite{Jeltes2007}.

\noindent Metastable helium experiments have been carried out by a
number of groups. In particular, the group of Westbrook and Aspect
at the Institut d'Optique (France) have already observed three-dimensional
collisions of metastable He{*} , using multichannel plate (MCP) detection
combined with a time-domain multiplexor to obtain both temporally
and spatially resolved quantum correlations of atomic arrival times\cite{Westbrooke2006,Perrin2007}.
Backward and forward quantum correlations were observed to be enhanced.
A similar experiment is also underway by Truscott\cite{Dall2007}
at The Australian National University.

\section{Many-body quantum dynamics}

The well-known difficulty with treating the dynamical quantum theory
of many-body systems, is that the Hilbert space --- the number of quantum
states involved --- can become exponentially complex. This is a subtle
point, as of course given an exact solution, a system in a pure state
is described by just one quantum state at all times. The problem is
that when one does not know the relevant exact quantum state, it is
necessary to expand in a basis of states.

\subsection{Exponential complexity}

As an example, consider $n$ atoms distributed among $m$ modes, with
$n\simeq m\simeq500,000$. The number of possible many-body number
states that could be involved is:

\begin{equation}
N_{s}=2^{2n}=2^{1,000,000}\end{equation}
Since this number means that there are more quantum states than atoms
in the universe, we conclude that even on the largest possible computer,
we can't diagonalize the Hamiltonian relative to this basis, in general!

\subsection{Traditional theoretical methods}

There are many traditional theoretical methods, which all have severe
drawbacks as first principles solutions:

\begin{enumerate}
\item Perturbation theory. A well established approach,  perturbation theory
can be applied in many ways, and a particularly sophisticated variation
is obtained by using perturbation theory on a path-integral formulation
of quantum dynamics. This approach diverges at strong couplings and
long times. 
\item Operator factorization. This approach neglects any quantum correlations.
Although many variations exist, these methods inherently involve uncontrolled
approximations, and are not applicable for strong correlations. 
\item Restricted Hilbert-spaces. Complementary to the idea of operator factorization
is the approximation of using a truncated Hilbert space, with an unknown
error due to the truncation. Examples in this general category include
the density matrix renormalization method\cite{Schollwock2005}. Similarly,
density functional theory\cite{Kohn1999} has unknown approximation
errors. 
\item Numerical diagonalisation. While exact in principle, this is intractable
for large particle numbers, unless the number of spatial modes is
severely restricted. 
\item Bethe ansatz solutions. Certain one-dimensional many-body problems
have exactly known eigenstates from the Bethe ansatz. These can be
 very useful in static cases. However, knowing  the eigenvaluesdoes
not necessarily solve the dynamical complexity problem. Exponentially
many eigenstates are still required to expand an arbitrary initial
state --- there are simply too many basis states for exact quantum dynamical
calculations. 
\item Quantum computers. Can quantum computers solve quantum dynamics? In
1982, Feynman proposed this approach. By even the most optimistic
predictions, hardware of practical use is still many years away, and
their range of application appears limited. 
\end{enumerate}
In summary, we see that while experimentalists have more sophisticated
tools than ever before, the theorist faces severe difficulties in
modeling these new experiments. It would clearly be useful to have
first-principles techniques that utilize existing computers.

\subsection{Classical phase space }

One of the most important and enduring ideas of Glauber\cite{Glauber:1963},
developed in parallel with an approach of Sudarshan\cite{Sudarshan:1963},
was the use of coherent states to generate quantum operator representations
for bosons. In some cases one can obtain expansions of the density
matrix using a probability $P(\vec{\alpha})$. For an $M$-mode bosonic
problem, we define:

\begin{equation}
\,\widehat{\rho}=\int P(\vec{\alpha})\left|\vec{\alpha}\right\rangle \!\!\left\langle \vec{\alpha}\right|d^{2M}\vec{\alpha}\,\,.\end{equation}
where $\left|\vec{\alpha}\right\rangle $ is an $M$-mode coherent
state, defined as a simultaneous eigenstate of the annihilation operators.
This approach maps quantum states into an essentially classical phase-space.
We note that there is a clear limitation here: the expansion is a
separable one, and therefore cannot describe entangled states.

The technique, of course, was highly successful in its applications
to the quantum theory of the laser, since a laser output state is
typically non-entangled. Different variations of this approach are
obtained by considering different operator orderings in the equivalence
relations between operator products and classical field products.
Many prominent physicists have developed and used phase-space distributions
for quantum systems, starting from Wigner\cite{Wigner1932a} and Husimi\cite{Husimi:1940},
with later developments due to Glauber, Sudarshan, Agarwal and Wolf\cite{Agarwal:1970,Agarwal:1970a},
Lax\cite{Lax:1966} and many others.

The problem, however, with interpreting these distributions as  probabilistic
mappings to a classical phase space is that these are fundamentally
incomplete. When used to calculate general quantum dynamical time-evolution,
either the distributions or the propagators can have negative values.
Even the Husimi Q-function, which is statically complete and always
positive, generally has no corresponding positive propagator. Hence,
no stochastic process is available for simulation purposes.

\subsection{Quantum phase space }

The problems of exponential complexity can be reduced --- though not
wholly eliminated --- by using a doubled phase-space expansion that
allows quantum superpositions and entanglement in the basis set. The
idea of dimension-doubling was also proposed by Glauber\cite{Glauber:1963}.
However, this by itself is not sufficient. It is also necessary to
have an appropriate differential mapping, which maps the operator
products that occur in a physical Hamiltonian, to positive definite
differential operators that have stochastic equivalences.

The first approach of this type was the positive-P representation\cite{Drummond1980a,Chaturvedi:1977a},
in which there are $2M$ complex coordinates, so that:

\begin{equation}
\,\widehat{\rho}=\int P(\vec{\alpha},\vec{\beta})\frac{\left.\left.\left|\vec{\beta}\right\rangle \right\langle \vec{\alpha}\right|}{\left\langle \vec{\alpha}\left|\vec{\beta}\right.\right\rangle }d^{2M}\vec{\alpha}d^{2M}\vec{\beta}\label{eq:posp}\end{equation}
 The resulting distributions are positive and obey a diffusion equation,
so that they can be effectively simulated using a stochastic process.

\subsection{Application}

Before turning to specific examples, we give the $+P$ equations for
a general bosonic system with two-body interactions. Such systems
are modeled by using nonlinear interactions on a lattice, together
with linear interactions coupling different sites, so that the quantum
Hamiltonian is:

\begin{equation}
\,\widehat{H}(\mathbf{a},\mathbf{a}^{\dagger})=\hbar\sum_{ij}\left[\omega_{ij}a_{i}^{\dagger}a_{j}+\frac{1}{2}\chi_{ij}:\widehat{n}_{i}\widehat{n}_{j}:\right]\,.\end{equation}
 Here $\,\omega_{ij}$ is a nonlocal linear coupling, which may correspond
to simple quantum diffusion of free particles, or else to inter-well
hopping in the case of a true lattice, while $\chi_{ij}$ is a nonlocal
nonlinear coupling. If $\chi_{ij}=\chi\delta_{ij}$, then one recovers
the usual local interaction lattice theory, applicable for ultracold
atoms under s-wave scattering. The boson number operator at each site
is: $\,\widehat{n}_{i}=a_{i}^{\dagger}a_{i}$, which has a stochastic
equivalent of $n_{i}=\beta_{i}^{*}\alpha_{i}$. Even though the Hamiltonian
has the appearance of modeling a lattice, the general approach also
holds for quantum fields with a momentum cut-off that equals the inverse
lattice spacing.

With the addition of nonlocal linear damping of $\kappa_{ij}$, the
simplest corresponding positive-P stochastic equations have the It\^o
form:

\begin{eqnarray}
\frac{\partial\alpha_{i}}{\partial t} & = & -\left(\kappa_{ij}+i\omega_{ij}\right)\alpha_{j}-\left(i\chi_{ij}n_{j}+b_{ik}\eta_{k}^{(a)}(t)\right)\alpha_{i}\nonumber \\
\frac{\partial\beta_{i}}{\partial t} & = & -\left(\kappa_{ij}+i\omega_{ij}\right)\beta_{j}-\left(i\chi_{ij}n_{j}^{*}+b_{ik}\eta_{k}^{(b)}(t)\right)\beta_{i}\end{eqnarray}
Here the noise matrix $\mathbf{b}$ is the solution to $b_{ik}b_{jk}=-i\chi_{ij}$,
and the noises are delta correlated, so that $\langle\eta_{i}^{(a)}(t)\eta_{j}^{(a')}(t')\rangle=\delta_{ij}\delta^{(a,a')}\delta(t-t')$. 

The earliest example of this technique was the prediction of quantum
squeezing of solitons in optical fibres\cite{Carter1987a}, using
the positive-P representation. In this case the relatively high occupations
of modes means that a truncated form of the Wigner representation
is also very useful. These predictions have been recently confirmed
to high accuracy. There are many other applications, including simulation
of evaporative cooling\cite{Drummond1999}, spin squeezing \cite{Poulsen2001b},
correlation dynamics in a uniform gas\cite{Deuar2006a}, the quantum
evolution of large numbers number of interacting atoms in a single
well\cite{Dowling2005}, the dynamics of atoms in a 1D trap\cite{Carusotto2001},
and molecular down-conversion to atoms\cite{Savage2006}.

\section{Examples}

In the remainder of this paper, we shall focus on some novel results.

\subsection{Direct quantum simulations of BEC formation}

A thorough treatment of the initial state of a quantum experiment
ideally should include a theory of state-preparation. While it is
certainly possible to use phase-space techniques starting from the
common assumption of a canonical ensemble at finite temperature, there
is a more fundamental question of interest in ultra-cold atom Bose-Einstein
condensation. Is the concept of a thermal equilibrium at finite temperature
always applicable in these experiments? This question arises because
the experiments are fundamentally non-equilibrium in nature, with
no external reservoir at a fixed temperature as in most condensed
matter experiments.

\begin{figure}
\begin{centering}\includegraphics[width=6cm]{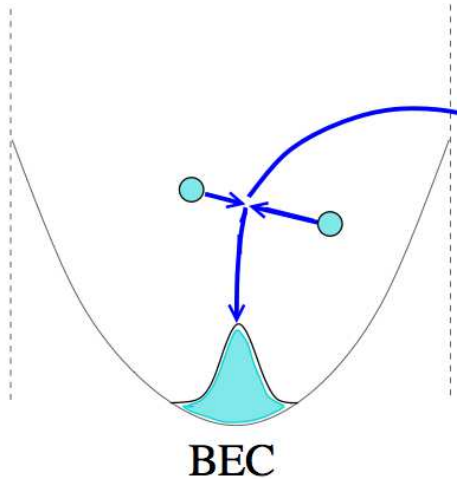} \par\end{centering}

\caption{diagram of the physics of evaporative cooling. Two atoms collide,
one losing energy and becoming condensed, while the other gains energy
and escapes from the trap.\label{fig:diagram-of-the}}
\end{figure}

The true state of an atom laser or BEC is the result of cooling through
evaporation. Therefore to try to answer this question, one must simulate
the actual evaporative cooling process that leads to condensate formation.
The process itself is depicted schematically in Fig (\ref{fig:diagram-of-the}).
Collisions of hot atoms lead to condensate formation together with
the escape of even hotter atoms from the trap, as there is an overall
energy conservation in the collisions.

In our simulations, we use a model identical to that used in\cite{Drummond1999}:
a small $(3+1)$D system with an initial $240$nK thermal distribution.
We suppose there are $N=10^{4}$ bosonic atoms of mass $m=1.5\times10^{-25}$kg,
initially confined to a box of side $L=10\mu$m. At $t=0^{+}$ a
smooth trapping potential is switched on. It is of the form\begin{equation}
V(\vec{x},t)=V_{0}(1-t/t_{0})\sum_{j=1}^{D}\sin^{2}\left(\frac{\pi x_{i}}{2L}\right)\end{equation}
 where $k_{B}V_{0}$ is similar to the initial temperature and $t_{0}$
is the length of the relaxation --- about 100ms. This potential is chosen
so that atoms can escape more readily as time evolves, which should
lead to a continual lowering in the average energy of the remaining
atoms. This cooling strategy also reduces the effective trap frequency
with time, leaving a cloud of untrapped atoms at the end of the process.
In the simulations, the lattice boundary is absorbing, so that atoms
reaching the edge of the simulation region are simply removed through
localized linear damping.

\subsubsection{Definition of COM temperature}

While it is widely accepted that the evaporative cooling strategy
leads to formation of a Bose-Einstein condensate, the question of
which mode is condensed is not so easily answered. Simulations indicate
that the condensate is in motion, with a center of mass effective
temperature that may both increase as well as decrease during evaporative
cooling. In other words, while the relative motion of particles is
being cooled, it is possible for the center-of-mass to become hotter,
since these degrees of freedom are largely decoupled.

We can directly apply the equipartition theorem to arrive at an estimate
of the COM temperature. We assume that the COM energy can be written\begin{equation}
E_{COM}\sim\frac{\langle|\vec{P}|^{2}\rangle}{2m\langle\hat{N}\rangle}+\frac{1}{2}m{(\omega}_{\textrm{eff}}{(t))}^{2}\frac{\langle|\hat{N}\vec{X}|^{2}\rangle}{\langle\hat{N}\rangle}\end{equation}
 where $\omega_{\textrm{eff}}(t)$ is obtained from $V(\vec{x},t)$
by assuming small deviations of the COM position from zero, in which
case $\sin^{2}x\simeq x^{2}$. We then estimate that the effective
temperature is given by\begin{equation}
T_{COM}\simeq\frac{E_{COM}}{Dk_{B}}\end{equation}
 where $D$ represents twice the number of COM degrees of freedom
(eg.~$6$ for a trapped 3D gas).

\subsubsection{Positive-P simulation results}

\begin{figure}

\begin{centering}\includegraphics[width=10cm,keepaspectratio]{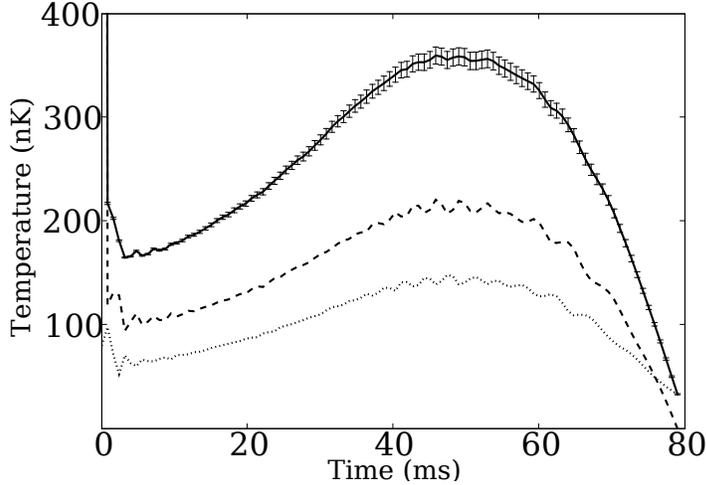} \par\end{centering}

\caption{Full 3D positive-P results for COM heating. The solid curve represents
the full COM temperature estimate, where the large initial spike is
due to the rapid switch-on of the trap. The dotted and dashed curves
represent the contributions of kinetic and potential energy respectively.\label{fig:Full-3D-positive-P}}
\end{figure}

Fig (\ref{fig:Full-3D-positive-P}) gives a full three-dimensional
simulation of this evaporative cooling scenario, focusing on the effective
temperature of the center-of-mass.

It is clear from these results that a strong evaporative heating of
the center-of-mass degree of freedom occurs simultaneously with the
evaporative formation of a Bose condensate. This can be understood
from physical arguments. The Bose enhancement of scattering into moving
modes randomly condenses the gas into a moving condensate. Here the
coherence properties of the gas means that all the condensed atoms
have the same velocity, which enhances the effective center-of-mass
temperature, even while the condensate forms. The final cooling of
the center-of-mass temperature is due to enhanced evaporative losses
of more rapidly moving condensates, as the trap walls are lowered.
We note here that these results are limited by the sampling error,
depicted by the error bars in the figure.

\subsection{Long-range or strong particle interactions}

The positive-P and related phase-space representations are also readily
applicable to strongly interacting situations where the range or shape
of the interparticle interaction cannot be ignored. Quantitative dynamics
for such systems have been very difficult to obtain apart from some
special systems.

Here, instead of pre-calculating scattering behaviour for two- (or
more- ) body collisions, we remain with the raw Hamiltonian that explicitly
gives an interparticle potential. This corresponds to the nonlocal
form of the lattice Hamiltonian given above. As a proof of principle,
we have calculated the dynamics of a small strongly interacting cloud
of cold bosons confined in a one-dimensional trap whose width is of
the same size as the range of the interparticle potential\cite{Deuar2005}.

At $t<0$, bosons with negligible interparticle interaction are prepared
in a harmonic trap with trapping potential $V^{{\rm ext}}(x)=\frac{1}{2}\ m\ \omega_{{\rm ho}}^{2}\ x^{2},$
which has a harmonic oscillator length $a_{{\rm ho}}=\sqrt{\hbar/m\omega_{{\rm ho}}}$.
Initially they are in the coherent zero-temperature ground state obtained
by solving the Gross-Pitaevskii mean field equations\cite{Dalfovo1999}.
The mean number of atoms in the trap in this example is $\bar{N}=10$
($\bar{N}=100$ was also simulated, albeit for a shorter time span).
At $t=0$, interparticle interactions are turned on across the system.
This kind of effect is most commonly induced in practice by utilizing
a Feshbach resonance. A {}``breathing'' of the atomic cloud is also
induced by switching the trap to a more confined harmonic potential
with double the trapping frequency. 

We model the interparticle interactions by a Gaussian interparticle
potential \begin{equation}
U(x)=\gamma\ (\hbar\,\omega_{{\rm ho}})\left(\frac{1}{\sigma_{U}\sqrt{2\pi}}\right)\exp\left[-\frac{1}{2}\left(\frac{x}{a_{{\rm ho}}\,\sigma_{U}}\right)^{2}\right].\end{equation}
 Here $\gamma$ is a dimensionless strength parameter, and $\sigma_{U}$
is the standard deviation of the potential's shape (in units of $a_{{\rm ho}}$). 

\begin{figure}
\begin{centering}\includegraphics[width=12cm]{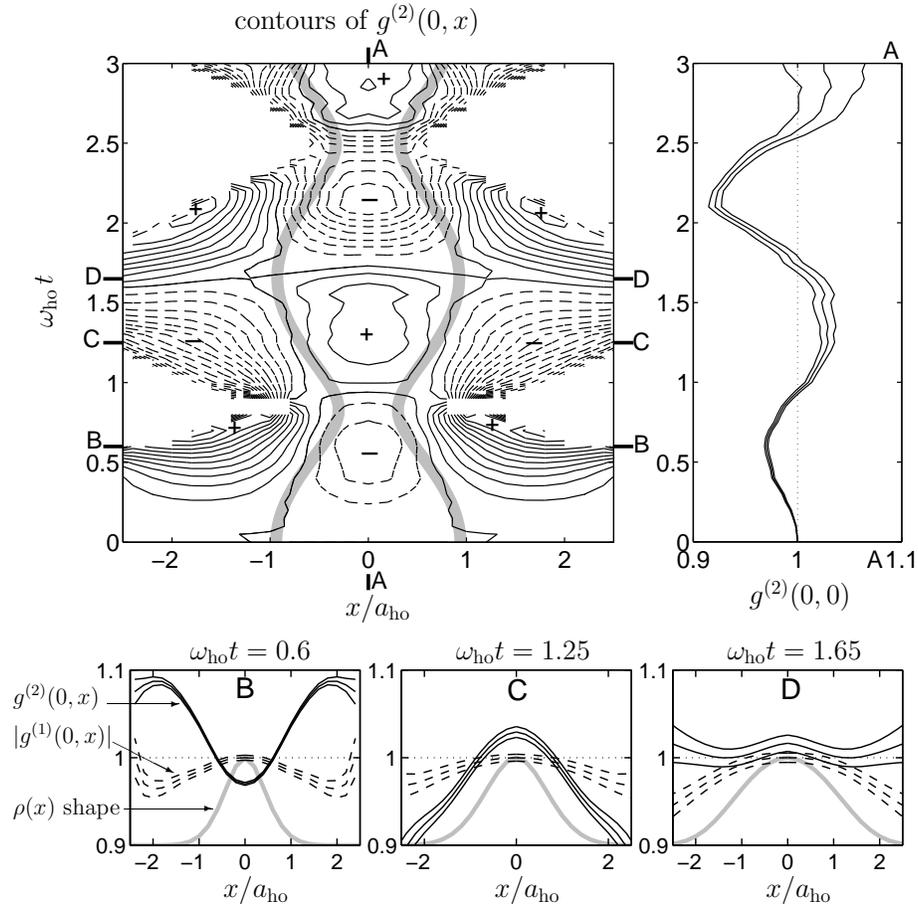} \par\end{centering}

\caption{\label{figbreath} Dynamics of correlations with long-range interactions. }
\end{figure}

For $\bar{N}=10$, $\gamma=0.4$, and $\sigma_{U}=1$, one obtains
the results shown in Figure~\ref{figbreath}. The simulation was
carried out on a $M=60$ lattice with $L=12a_{{\rm ho}}$, and $\mc{S}=10^{4}$
trajectories. We calculate the correlations between particles in the
middle of the cloud and in the wings. Such ranged two-body correlations
give insight into what behaviour may be expected to be typical during
a single experimental run. The first-order correlation function \begin{equation}
g^{(1)}(0,x)=\frac{\langle\dagop{\Psi}(0)\op{\Psi}(x)\rangle}{\sqrt{\rho(0)\,\rho(x)}},\end{equation}
 describes coherence between particles, where the density is $\rho(x)=\langle\dagop{\Psi}(x)\op{\Psi}(x)\rangle$.
The second-order (number) correlation function \begin{equation}
g^{(2)}(0,x)=\frac{\langle\dagop{\Psi}(0)\dagop{\Psi}(x)\op{\Psi}(0)\op{\Psi}(x)\rangle}{\rho(0)\,\rho(x)}\end{equation}
 describes number correlations. That is, when $g^{(2)}(0,x)>1$, the
likelihood of observing a pair of particles with one in the centre
of the trap ({}``0''), and one at $x$ is increased with respect
to the baseline case where the occupation at these points is uncorrelated
and given purely by the local densities. In the largest panel, contours
of the number correlation function $g^{(2)}$ are plotted, with solid
contours indicating $g^{(2)}(0,x)\ge1$, dashed contours indicating
$g^{(2)}(0,x)<1$. Contour spacing is 0.01. The thick grey lines indicate
the rms width of the cloud density $\rho(x)$. In panels A--D, triple
lines indicate one-sigma uncertainty, solid lines show $g^{(2)}$,
dashed lines show the coherence $|g^{(1)}|$, and solid thick grey
lines the density $\rho(x)$.

Notable features seen include the following.

\begin{itemize}
\item In the middle of the trap, there is an oscillation between bunching
($g^{(2)}(0,0)>1$) and antibunching ($g^{(2)}(0,0)<1$). 
\item This oscillation is \textsl{out of phase} (by approximately $\pi/2$)
with respect to the breathing motion of the gas cloud, and the behaviour
of pair correlations is quite counter-intuitive. In particular,

\begin{itemize}
\item When the particle cloud is contracting, \textsl{anti}bunching appears
at the center of the trap despite a net motion of particles into this
region, while simultaneously there is an enhanced likelihood of pairs
of atoms with one in the outer region of the cloud and one in the
center. 
\item During expansion, on the other hand, the particles in the center of
the trap tend to appear there in pairs despite the net flow of particles
out of this region, while pairs of particles with one in the tails,
one in the center are suppressed. 
\end{itemize}
\item The oscillations of $g^{(2)}(0,0)$ (at the center of the trap) become
more pronounced with time. This may indicate a resonance between the
breathing and the repulsion, although it is also possible that this
is a transient initial effect. 
\item Coherence between the center of the trap and outlying regions of the
cloud deteriorates as time proceeds. 
\end{itemize}

\section{General phase-space representations}

\begin{figure}
\begin{centering}\includegraphics[width=6cm]{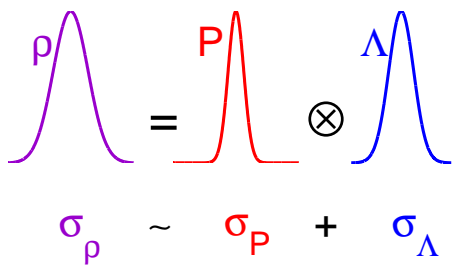} \par\end{centering}

\caption{Schematic diagram of how density matrix operator variances are composed
of a distribution variance and a basis variance.\label{fig:Schematic-diagram-of}}
\end{figure}

Finally, we show how these coherence-theory methods can be generalised
to incorporate other kinds of correlation into the basis. Most generally,
the phase-space approach can be defined as an expansion of the density
matrix $\widehat{\rho}$ , using nonorthogonal operators $\,\widehat{\Lambda}(\overrightarrow{\lambda})$,
such that:

\begin{equation}
\,\widehat{\rho}=\int P(\overrightarrow{\lambda})\widehat{\Lambda}(\overrightarrow{\lambda})d\overrightarrow{\lambda}\end{equation}
 Provided suitable differential identities exist, and that it is possible
to integrate by parts, quantum dynamics is transformed to a set of
stochastic trajectories in the generalized phase-space variable $\overrightarrow{\lambda}$.
A different basis choice leads to a different representation. Thus,
for example, in the positive P-representation, $\widehat{\Lambda}(\overrightarrow{\lambda})$
is the off-diagonal coherent-state projector apearing in equation
Eq.\ (\ref{eq:posp}).There are a number of clear trade-offs, in
that a variance can be transferred from the distribution to the basis,
in order to obtain reducing trajectory spread, leading to lower sampling
error. This is shown schematically in Fig (\ref{fig:Schematic-diagram-of}).

\subsection{General $M$-mode Gaussian operator}

Many possible basis sets can be used. As a generic form applicable
to both fermionic and bosonic cases, we may consider a Gaussian operator
basis, defined here as the normally ordered exponential of a quadratic
form in the $2M$-vector mode operator $\delta\underline{\widehat{a}}=(\widehat{\mathbf{a}},\widehat{\mathbf{a}}^{\dagger})-\underline{\alpha}$
, where $\underline{\alpha}$ is a c-vector and $\widehat{\mathbf{a}}$
is the vector of annihilation operators. The bosonic kernel\cite{Corney:2003},
with a similar result for fermions\cite{Corney2006b}, is:\begin{equation}
\,\widehat{\Lambda}(\overrightarrow{\lambda})=\frac{\Omega}{\sqrt{\left|\underline{\underline{\sigma}}\right|}}:\exp\left[-\delta\underline{\widehat{a}}^{\dagger}\underline{\underline{\sigma}}^{-1}\delta\underline{\widehat{a}}/2\right]:\,\,.\end{equation}

Here the `quantum phase space' is extended even further, to the vector:
$\overrightarrow{\lambda}=(\Omega,\underline{\alpha},\underline{\underline{\sigma}})$.
This now includes the covariance $\underline{\underline{\sigma}}$,
which can be readily parametrized in terms of normal and anomalous
Green's functions, denoted $\mathbf{n}$ and $\mathbf{m}$ respectively:

\begin{equation}
\,\underline{\underline{\sigma}}=\left[\begin{array}{cc}
\mathbf{I}+\mathbf{n} & \mathbf{m}\\
\mathbf{m}^{+} & \mathbf{I}+\mathbf{n}^{T}\end{array}\right]\,\,.\end{equation}
When $\mathbf{n=m=0}$, the representation reduces to that of the
positive-P. However with these new parameters, the representation
is complete (for number-conserving systems) when the coherent amplitudes
are zero, thus allowing for representation of Fermions.

In summary, the Gaussian representation phase space is $\,\overrightarrow{\lambda}=(\Omega,\bm\alpha,\bm\beta,\mathbf{n},\mathbf{m},\mathbf{m}^{+})$,
where  $\Omega$ is a weight factor, $\alpha,\beta$ are (for bosons)
coherent  amplitudes, $\mathbf{n}$ is number correlations and $\mathbf{m},\mathbf{m}^{+}$
 are squeezing correlations.

\subsection{Weighted stochastic gauge equations}

The use of phase-space methods has a fundamental philosophy of attempting
to transform \textbf{hard} quantum problems into \textbf{tractable}
stochastic equations. However, there are several ways to do this,
due to the overcompleteness of the basis set. For a basis set that
is analytic in the phase-space variables, one can show that, provided
partial integration is possible, one can obtain an equivalence class
of stochastic equations. These include an arbitrary `stochastic gauge'
function $\mathbf{g}$\cite{Deuar:2002}, and have the generic structure:
\begin{eqnarray}
d\Omega/\partial t & = & \Omega\left[U+\,\mathbf{g}\,\cdot\bm\zeta\right]\nonumber \\
d\bm\alpha/\partial t & = & \mathbf{A}+\mathbf{B}(\bm\zeta-\,\mathbf{g})\,.\end{eqnarray}

In principle, the  Gaussian basis allows a wide range of fermionic
and bosonic systems to be simulated from first principles\cite{Corney2004a}.
Nevertheless, there are unsolved problems that remain (see e.g. \cite{Assaad2005a}).
The chief issue is the sampling error, since typically many stochastic
trajectories are needed to control growing sampling errors, which
eventually become too large for useful results. The sampling error
can be improved through careful choice of the gauge function \textbf{$\mathbf{g}$},
which is a function chosen to stabilize trajectories, as well as the
basis set itself, and the details of the simulation method.

\section{Summary}

In summary, we have described areas of recent development --- one experimental
and one theoretical --- in many-body quantum physics that can be traced
from the coherence theory that originated in quantum optics. 

Experiments in ultracold atoms are increasingly able to probe the
quantum correlations that arise from the many-body nature of the system.
We have here been able to describe just two --- those involving dynamical
collisions of dense clouds --- but there are many others. 

In parallel, coherence theory has lead to a range of powerful phase-space
techniques to simulate many-body quantum dynamics. We have discussed
the positive-P approach and its generalisation, the generalised Gaussian
method.  These methods give rise to representations for bosons \textbf{and}
fermions, and can deal with either local or nonlocal interactions.
One can readily perform three-dimensional lattice simulations, with
up to up to $10^{23}$ particles, and $10^{6}$ modes. The fact that
sampling errors increase with time is a serious limitation, however.

Novel physical effects found in the simulations include evaporative
heating (not cooling) of center of mass temperatures, and an unexpected
atomic anti-bunching effect during compression of a condensate with
long-range interactions.

There are many interesting challenges and new quantum physics to be
investigated with these approaches. As well as ultra-cold atoms, other
complex systems may be investigated, ranging from nanotechnology,
through to biochemistry and genetics \cite{Drummond:2004}.

\section*{Acknowledgements}
 We acknowledge funding from the Australian
Research Council Centre of Excellence program, and useful discussions
with K.V. Kheruntsyan. This work was also supported financially by
the NWO as part of the FOM quantum gases project.

\bibliographystyle{unsrt}
\bibliography{Phasespacexv2}

\end{document}